\documentclass[12pt,oneside,english]{amsart}
\usepackage[T1]{fontenc}
\usepackage[latin9]{inputenc}
\usepackage{geometry}
\usepackage{enumitem}
\usepackage{mathrsfs}
\usepackage{amstext}
\usepackage{amsthm}
\usepackage{amssymb}
\usepackage{float}
\usepackage{textcomp} 

\usepackage[svgnames,table]{xcolor}
\usepackage[bookmarksnumbered=true]{hyperref} 
\hypersetup{
	colorlinks = true,
	linkcolor = Blue,
	anchorcolor = blue,
	citecolor = Green,
	filecolor = blue,
	urlcolor = FireBrick
}
\usepackage{bbm}
\usepackage{comment}

\usepackage[nameinlink]{cleveref}
\Crefname{assumption}{Assumption}{Assumptions}
\usepackage{orcidlink}

\makeatletter
\numberwithin{equation}{section}
\numberwithin{figure}{section}
\numberwithin{table}{section}
\theoremstyle{definition}
\newtheorem{theorem}{Theorem}[section]
\theoremstyle{definition}

\theoremstyle{definition}
\newtheorem{remark}[theorem]{Remark}
\theoremstyle{definition}

\theoremstyle{definition}

\theoremstyle{definition}

\newlist{casenv}{enumerate}{4}
\setlist[casenv]{leftmargin=*,align=left,widest={iiii}}
\setlist[casenv,1]{label={{\itshape\ \casename} \arabic*.},ref=\arabic*}
\setlist[casenv,2]{label={{\itshape\ \casename} \roman*.},ref=\roman*}
\setlist[casenv,3]{label={{\itshape\ \casename\ \alph*.}},ref=\alph*}
\setlist[casenv,4]{label={{\itshape\ \casename} \arabic*.},ref=\arabic*}
\setitemize{leftmargin=*, align=left}
\providecommand{\casename}{Case}

\newtheorem*{remark*}{Remark}
\newtheorem*{example*}{Example}

\usepackage{babel}

\DeclareRobustCommand{\SkipTocEntry}[5]{}

\newcommand{\mR}{\mathbb{R}}   
\newcommand{\mC}{\mathbb{C}}   
\newcommand{\mN}{\mathbb{N}}   

\newcommand{\bfi}{\mathbf{i}} 
\newcommand{\rmd}{\mathrm{d}} 
\newcommand{\mS}{\mathcal{S}} 

\newcommand{\abs}[1]{\lvert #1 \rvert}  
\newcommand{\norm}[1]{\lVert #1 \rVert}  

\newcommand{\ehat}{\,\widehat{\rule{0pt}{6pt}}\,}

\newcommand{\tre}{\textcolor{red}}

\DeclareRobustCommand{\SkipTocEntry}[5]{}

\makeatother

\begin{document}
\title[FFT-ConvAE Model]{A Speech Enhancement Method Using Fast Fourier Transform and Convolutional Autoencoder}

\author[P.-Y. Kow]{Pu-Yun Kow\,\orcidlink{0000-0001-5718-9316}}
\address{Department of Artificial Intelligence, Tamkang University, New Taipei City 251301, Taiwan.} 
\email{169016@o365.tku.edu.tw}

\author[P.-Z. Kow]{Pu-Zhao Kow\,\orcidlink{0000-0002-2990-3591}}
\address{Department of Mathematical Sciences, National Chengchi University, Taipei 116, Taiwan.}
\email{pzkow@g.nccu.edu.tw}

\begin{abstract}
\begin{sloppypar}

This paper addresses the reconstruction of audio signals from degraded measurements. We propose a lightweight model that combines the discrete Fourier transform with a Convolutional Autoencoder (FFT-ConvAE), which enabled our team to achieve second place in the Helsinki Speech Challenge 2024. Our results, together with those of other teams, demonstrate the potential of neural-network-free approaches for effective speech signal reconstruction. 

\end{sloppypar}
\end{abstract}

\subjclass[2020]{68T07, 68T10, 68T20, 35R25, 35R30}
\keywords{inverse problem, Fourier transform, Convolutional-based Autoencoder (ConvAE), convolutional neural network (CNN), artificial neural network (ANN), Artificial Intelligent (AI)}

\maketitle
\tableofcontents{}

\begin{sloppypar}

\addtocontents{toc}{\SkipTocEntry}
\subsection*{Acknowledgments} 

This study is supported by the National Science and Technology Council of Taiwan, NSTC 112-2115-M-004-004-MY3. 


\section{Introduction\label{sec:Introduction}}

In this paper, we address the problem of reconstructing audio signals from noisy measurements, focusing on speech enhancement with the goal of improving the quality of signals degraded by adverse acoustic channel conditions. This problem has been studied extensively with a wide range of methods. Rather than providing a detailed literature review here, we direct the reader to the survey papers \cite{SpeechProcessingReview2023,FrequencyDomainMonauralSpeechEnhancement}. Specifically, we evaluate the performance of a speech enhancement method on the dataset and tasks of the Helsinki Speech Challenge 2024 (HSC2024 \cite{HSC2024}), where our team achieved second place.

The main challenge of this problem is processing a large number of samples while maintaining a low real-time factor (RTF), defined as the processing time divided by the audio length. In other words, high-dimensional data must be processed using lightweight models. According to the challenge rules \cite[Section~5.2]{HSC2024}, the average RTF must not exceed 3, meaning that no more than 3 seconds may be used to process each second of audio. Participants are therefore encouraged to develop effective, lightweight algorithms.

In this work, we address these constraints by applying an algorithm that combines the Fast Fourier Transform (FFT) with a Convolutional Autoencoder (ConvAE) model across all 12 levels.

\section{Speech enchancement as an inverse problem\label{sec:audio-IP}} 

The general speech enhancement problem can be formulated as recovering the clean speech signal $\{x^{\rm clean}(t)\}_{t\in(0,t_{*})}$ from a recorded signal $\{x^{\rm deg}(t)\}_{t\in(0,t_{*})}$, which is typically degraded. This can be expressed as 
\begin{subequations}\label{eq:continuous-model} 
\begin{equation}
x^{\rm deg}(t) = \mathscr{A}(x^{\rm clean})(t) + u(t) + w(t) \quad \text{for all $t\in (0,t_{*})$} 
\end{equation}
where $\mathscr{A}$ is a (possibly nonlinear) filter, $u(t)$ is a non-stationary interfering signal, and $w(t)$ is stationary noise. According to \cite{HSC2024}, recording systems can often be modeled as Linear Time-Invariant (LTI), in which case the general speech enhancement problem reduces to a deconvolution problem: 
\begin{equation}
x^{\rm deg}(t) = (k*x^{\rm clean})(t) + w(t) \quad \text{for all $t\in (0,t_{*})$} 
\end{equation}
\end{subequations}
where $*$ denotes convolution and $k(t)$ is the impulse response of the system. 

In real-world audio processing, signals are typically recorded at a sample rate of $f_{s}$, i.e., $f_{s}$ samples per second, with each sample represented as a floating-point number between $-1$ and $1$. For storage efficiency, the audio is usually stored in 16-bit integer format (16-bit PCM) by multiplying each sample by 32,767 and rounding to the nearest integer\footnote{There are exactly $2^{16} = 65,536$ integers ranging from $-32,767$ to $32,767$, which explains the term ``16-bit''.}), as also requested by the organizers \cite[Section~5.4]{HSC2024}. In this way, each audio signal can be represented as an integer-valued vector $\mathbf{x} = (x_1, \dots, x_\ell)$. The length of an audio signal is defined as ${\rm length}(\mathbf{x}) := \ell$. 
The \emph{Nyquist-Shannon criterion} \cite{Shannon1949} provided a condition under which a discretized signal $\mathbf{x}$ can approximate its continuous-time counterpart $\{x(t)\}_{t\in(0,t_{*})}$. Specifically, if the signal contains no frequency components above $B$ Hz, then a nearly perfect reconstruction is guaranteed provided that $B < f_{s}/2$. 

\begin{remark}\label{rem:bandwidth}
In our case, the sample rate of $f_{s}=16\,{\rm kHz}$, as requested by the organizers \cite[Section~5.4]{HSC2024}, captures frequencies up to $8\,{\rm kHz}$, which is sufficient for speech recording while reducing computational load compared to the standard rate of $44.1\,{\rm kHz}$, which captures frequencies up to $22,050\,{\rm Hz}$ and covers the human hearing range of approximately $20\,{\rm Hz}$ to $20\,{\rm kHz}$.
\end{remark}

Accordingly, the discrete version of the speech enhancement problem can be formulated as recovering the clean signal $\mathbf{x}^{\rm clean} \in \mR^{m}$ from a degraded recording $\mathbf{x}^{\rm deg} \in \mR^{m}$. This can be written as 
\begin{subequations}\label{eq:discrete-model} 
\begin{equation}
\mathbf{x}^{\rm deg} = \widetilde{\mathscr{A}}(\mathbf{x}^{\rm clean}) + \mathbf{u} + \mathbf{w} 
\end{equation}
where $\widetilde{\mathscr{A}}$ is a (possibly nonlinear) filter, $\mathbf{u}$ is a non-stationary interfering signal, and $\mathbf{w}$ is stationary noise. In the common case of a LTI system, speech enhancement problem reduces to a deconvolution problem: 
\begin{equation}
\mathbf{x}^{\rm deg} = \mathbf{k}*\mathbf{x}^{\rm clean} + \mathbf{w} 
\end{equation}
\end{subequations} 
where $*$ denotes discrete convolution and $\mathbf{k}$ is the discrete impulse response of the system.

\section{Methodology\label{sec:Methodology}}

Given a clean audio signal $\mathbf{x}^{\rm clean} \in \mR^{m}$ of length $m\in\mN$ and a degraded audio signal $\mathbf{x}^{\rm deg} \in \mR^{m}$, our goal is to construct an approximation $\mathbf{x}^{\rm approx} \in \mR^{m}$ of $\mathbf{x}^{\rm clean}$. Using the Fast Fourier Transform (FFT), we compute their discrete Fourier transforms $\widehat{\mathbf{x}}^{\rm clean} \in \mC^{m}$ and $\widehat{\mathbf{x}}^{\rm deg} \in \mC^{m}$, defined by 
\begin{equation*}
\hat{x}_{k} = \sum_{m=0}^{n-1} x_{m} \exp \left( -2\pi\bfi\frac{mk}{n} \right) \quad \text{for $k=1,\cdots,m$}, 
\end{equation*}
where $\bfi$ denotes the imaginary unit, and $\exp$ is the complex exponential given by Euler's formula. It therefore suffices to construct an approximation $\widehat{\mathbf{x}}^{\rm approx} \in \mC^{m}$ of $\widehat{\mathbf{x}}^{\rm clean}$, since its inverse Fourier transform yields the desired approximation $\mathbf{x}^{\rm approx}$.

We consider an approximator of the form
\begin{equation}
\widehat{\mathbf{x}}^{\rm approx} = (\hat{x}_{1}^{\rm approx},\cdots,\hat{x}_{m}^{\rm approx}) \quad \text{with} \quad \hat{x}_{j}^{\rm approx} = A_{j} \frac{ \hat{x}_{j}^{\rm deg} }{\abs{ \hat{x}_{j}^{\rm deg} }} \label{eq:approximator}
\end{equation}
for some $A_{j}\in\mR$. Specifically, we use $\left( \abs{ \hat{x}_{1}^{\rm deg}}  ,\cdots , \abs{ \hat{x}_{m}^{\rm deg}} \right)$ to construct an estimator $\mathbf{A}=(A_{1},\cdots,A_{m})$ of $\left( \abs{\hat{x}_{1}^{\rm clean}} ,\cdots , \abs{\hat{x}_{m}^{\rm clean}} \right)$. Our approach can be viewed as a variant of the spectral gain method, which is classified as a classical technique in \cite{FrequencyDomainMonauralSpeechEnhancement}. We employ a Convolutional Autoencoder (ConvAE) to construct an estimator $\mathbf{z}$ based on all data within each task and level, making it specific to that task and level.

Convolutional Neural Networks (CNNs) are widely used in deep learning due to their strong ability to extract nonlinear features from input data (see, e.g., \cite{KLSYC2022ConvLSTM-CNN-BP}). Autoencoders (AEs) are an important AI architecture capable of denoising both image and time-series datasets (see, e.g., \cite{GLLSMVH2019AE,WCWW2020AE,XMY2016AE}) and handling high-dimensional data thanks to their data compression capability. The Convolutional Autoencoder (ConvAE) builds on the AE architecture by replacing hidden layers with CNN layers (see, e.g., \cite{CSTK2018ConvAE,KLSCC2024,WHKCC2023AE-CNN-BP}). ConvAEs have been successfully applied in various fields, for example, \cite{KLSCC2024} demonstrates their effectiveness in forecasting watershed groundwater levels.

To provide a clearer understanding, we first describe the fully connected Autoencoder (AE), which consists of two parts: the \emph{encoder} and the \emph{decoder}. 
Suppose the encoder and decoder have $\ell_{*}$ and $L_{*}$ hidden layers, respectively. Let $\phi : (0,\infty) \rightarrow \mR$ be an injective function with inverse $\phi^{-1} : \phi((0,\infty))\rightarrow (0,\infty)$ to be specified separately for each level and task. The encoder starts with input
\begin{equation*}
m_{0}=m ,\quad \mathbf{y}^{0} := \left( \phi\left(\abs{ \hat{x}_{1}^{\rm deg} }\right) ,\cdots , \phi\left(\abs{ \hat{x}_{m}^{\rm deg} }\right) \right) \quad \text{and activation functions $\left\{f^{\ell}\right\}_{\ell=1}^{\ell_{*}}$.} 
\end{equation*}
The state vector $\mathbf{y}^{\ell}\in\mR^{m_{\ell}}$ of the $\ell^{\rm th}$ hidden layer is given by 
\begin{equation*}
\mathbf{y}^{\ell} = f^{\ell}\left( \mathsf{w}^{\ell}\mathbf{y}^{\ell-1} + \mathbf{a}^{\ell} \right) \quad \text{for all $\ell=1,\cdots,\ell_{*}$} 
\end{equation*}
where $\mathsf{w}^{\ell}\in \mR^{m_{\ell}\times m_{\ell-1}}$ and $\mathbf{a}^{\ell}\in\mR^{m_{\ell}}$ are \emph{weights}. Here, $m_{\ell}\in\mN$ is the number of neurons in the $\ell^{\rm th}$ hidden layer. Expanding $\mathbf{y}^{\ell}=(y_{1}^{\ell},\cdots,y_{m_{\ell}}^{\ell})\in\mR^{m_{\ell}}$, each $y_{j}^{\ell}\in\mR$ represents the state of the $j^{\rm th}$ neuron. The final vector $\widetilde{\mathbf{y}}^{\ell_{*}}=(y_{1}^{\ell_{*}},\cdots,y_{m_{\ell_{*}}}^{\ell_{*}})$ serves as the encoded representation. 
The decoder begins with input 
\begin{equation*}
n_{0}=m_{\ell_{*}}, \quad \mathbf{z}^{0} := \widetilde{\mathbf{y}}^{\ell_{*}}=(y_{1}^{\ell_{*}},\cdots,y_{m_{\ell_{*}}}^{\ell_{*}}) \quad \text{and activation functions $\left\{g^{L}\right\}_{L=1}^{L_{*}}$.} 
\end{equation*}
The state vector $\mathbf{z}^{\ell}\in\mR^{n_{L}}$ of the $L^{\rm th}$ hidden layer is given by 
\begin{equation*}
\mathbf{z}^{L} = g^{L} \left( \mathsf{W}^{L}\mathbf{z}^{L-1} + \mathbf{b}^{\ell} \right) \quad \text{for all $L=1,\cdots,L_{*}$,} 
\end{equation*}
where $\mathsf{W}^{L}\in \mR^{n_{L}\times n_{L-1}}$ and $\mathbf{b}^{L}\in\mR^{n_{L}}$ are \emph{weights}. Finally, the output of the AE is 
\begin{equation*}
\mathbf{z}^{L_{*}} = \left(z_{1}^{L_{*}},\cdots,z_{L_{*}}^{L_{*}}\right), 
\end{equation*}
which is the decoded data. In our case, we set $L_{*}=m$ (as well as $\ell_{*}<m$) and construct the approximator \eqref{eq:approximator} with 
\begin{equation*}
A_{j} = \phi^{-1} \left( \Re z_{j}^{m} \right) \quad \text{for all $j=1,\cdots,m$,} 
\end{equation*}
where $\Re z_{j}^{m}$ denotes the real part of the complex number $z_{j}^{m}$.

\begin{figure}[H]
\centering
\includegraphics[width=.8\linewidth]{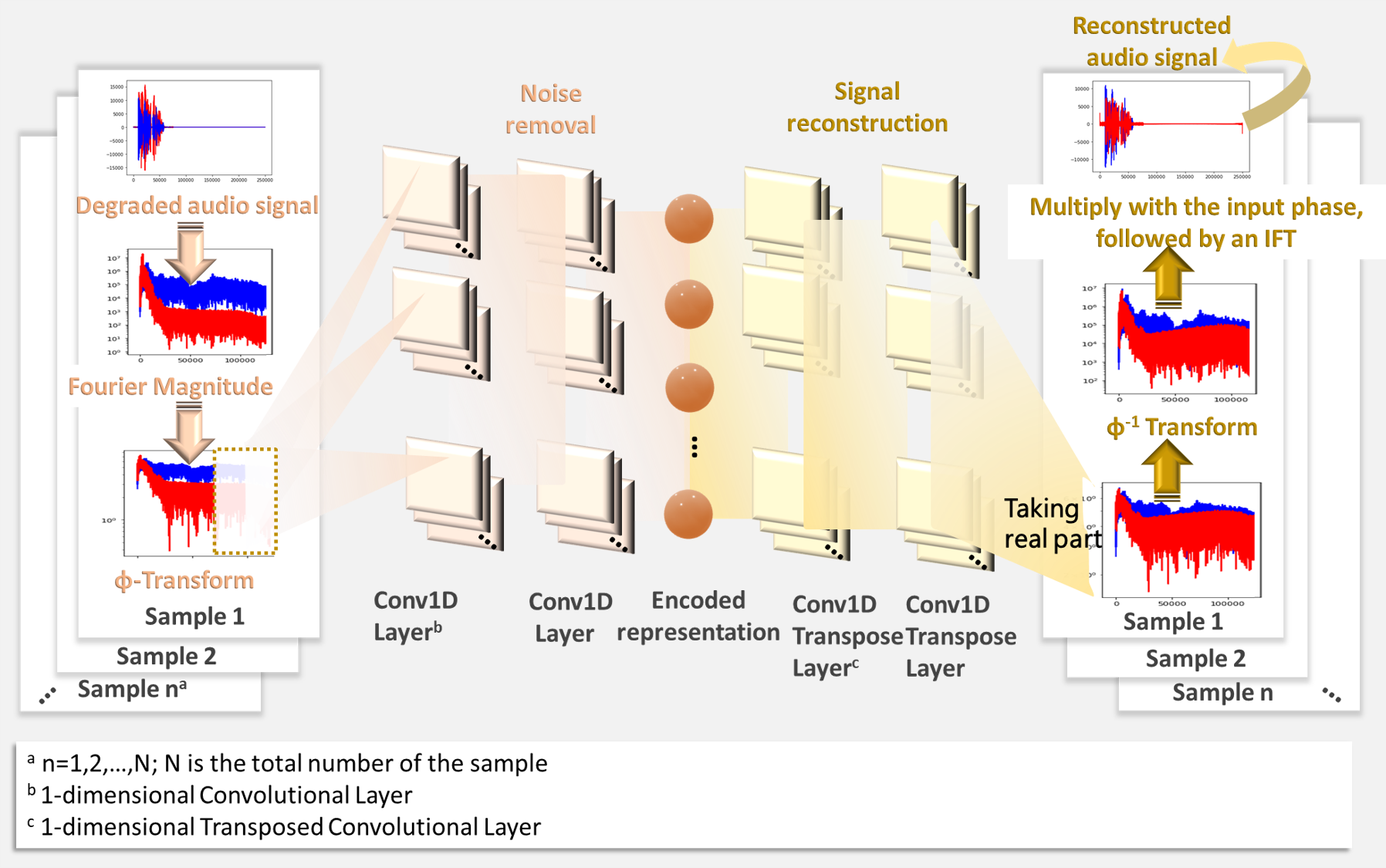}
\caption{Model architecture of the FFT-ConvAE Model} \label{fig:FFT-ConvAE}
\end{figure}

Mathematically, the Convolutional Autoencoder (ConvAE) can be seen as a special case of the fully connected AE. In this case, the weight matrix $\mathsf{w}^{\ell}\in \mR^{m_{\ell}\times m_{\ell-1}}$ in the $\ell^{\rm th}$ encoder layer takes the form 
\begin{equation*}
\mathsf{w}^{\ell}\mathbf{y} = \mathsf{k}^{\ell}*\mathbf{y} \quad \text{for all $\mathbf{y} \in \mR^{m_{\ell}}$}
\end{equation*}
where $\mathsf{k}^{\ell}$ is a convolution kernel and appropriate zero-padding is applied. Since $\mathsf{k}^{\ell}$ is a vector, this corresponds to a one-dimensional convolution layer (Conv1D layer). Similarly, the weight matrix $\mathsf{W}^{L}\in \mR^{n_{L}\times n_{L-1}}$ in the $L^{\rm th}$ decoder layer takes the form 
\begin{equation*}
(\mathsf{W}^{L})^{\intercal} \xi = \mathsf{K}^{L}*\xi \quad \text{for all $\xi \in \mR^{n_{L}}$}
\end{equation*}
where $\mathsf{K}^{L}$ is a convolution kernel and appropriate zero-padding is applied. Since $\mathsf{K}^{L}$ is a vector, this corresponds to a one-dimensional transposed convolution layer (Conv1D Transpose Layer). The architecture of our method, referred to as the FFT-ConvAE model, is shown in \Cref{fig:FFT-ConvAE}.

\section{Overview of the HSC2024\label{sec:dataset-description}} 

The dataset contains speech signals distorted by filtering and/or reverberation across 12 real-world-like setups. Specifically, seven filtering experiments (Levels 1--7) form Task~1, three reverberation experiments (Levels 1--3) form Task~2, and two experiments combining filtering and reverberation (Levels 1 and 2) constitute Task~3. Participants were tasked with developing algorithms to reconstruct unseen signals from the degraded data. The training datasets for the HSC2024 are available in the Zenodo repository \cite{HSC2024data}, and the test datasets are also provided there after the official results are published.

\subsection*{Task~1 (T1)}
\addtocontents{toc}{\SkipTocEntry}

This task involved 4,266 speech samples, played through a loudspeaker and captured by a microphone positioned at the opposite ends of soundproof tubes. A foam layer was inserted between them to mimic a low-pass filtering effect. The task is divided into seven levels (T1L1--T1L7), where each higher level corresponds to thicker layers of foam and additional materials, thereby making the filtering increasingly ill-posed.

\subsection*{Task~2 (T2)} 
\addtocontents{toc}{\SkipTocEntry}

In this setup, the loudspeaker and microphone were placed in a long enclosed hallway, producing 899 reverberant recordings. The task includes three levels (T2L1--T2L3), where the distance between the loudspeaker and microphone varies from 1 meter up to 10 meters.

\subsection*{Task~3 (T3)}
\addtocontents{toc}{\SkipTocEntry}
 
For this task, recordings from T1L2 and T1L4 were reused in the acoustic environment of T2L2 and T2L3, yielding two levels (T3L1 and T3L2). These recordings are simultaneously affected by high-frequency attenuation (from the filtering) and reverberation (from the hallway).

\subsection*{Evaluation} 
\addtocontents{toc}{\SkipTocEntry}

The organizers used Mozilla DeepSpeech\footnote{\url{https://github.com/mozilla/DeepSpeech}} to recognize speech from input \texttt{.wav} files, producing \texttt{.txt} transcripts, and compared the participating teams' results in terms of the character error rate (CER). Using a speech recognition model such as DeepSpeech to measure CER is uncommon in speech applications, as also noted by the organizers in \cite[Section~2.2]{HSC2024}. CER is defined as the ratio of incorrect or missing characters to the total number of characters in the reference text and is computed using \texttt{evaluate.py}. CER ranges from $0$ (all characters correct) to $1$ (all characters incorrect).

\section{Parameter Setting and Training\label{sec:training}}

We train separate models for each task and level, using the clean and degraded samples provided by the organizers. The model employs a symmetric encoder-decoder structure with only linear activations (i.e., no ReLU or other nonlinear functions), implemented by setting $f^{\ell}\equiv \rm Id$ and $g^{L}\equiv \rm Id$ across all tasks and levels. The encoder comprises three stacked Conv1D layers with decreasing filter sizes (64, 32, 16) and kernel sizes of 8 and 4, compressing the input into a lower-dimensional latent representation. Each layer is followed by batch normalization to stabilize and speed up training. The decoder mirrors this design with Conv1DTranspose layers (32, 64, 1) and kernel sizes of 8, reconstructing the original input (see \Cref{tab:model-param}). The model is trained end-to-end with the Adam optimizer at a learning rate of 0.001 for up to 100 epochs. Early stopping with a patience of 10 is used to halt training if the validation loss does not improve for 10 consecutive epochs, preventing overfitting and ensuring efficient convergence.

\begin{table}[H] 
\rowcolors{2}{lightgray!50}{white}
\renewcommand{\arraystretch}{1.5}  \centering
\begin{tabular}{c|c|c|c|c}
layer type & filters & kernel size & optimizer & loss function \\
\hline\hline
Conv1D & 64 & 8 & Adam & MSE \\ 
BatchNormalization & -- & -- & -- & -- \\ 
Conv1D & 32 & 8 & Adam & MSE \\ 
BatchNormalization & -- & -- & -- & -- \\ 
Conv1D & 16 & 4 & Adam & MSE \\ 
Conv1D Transpose & 32 & 8 & Adam & MSE \\ 
BatchNormalization & -- & -- & -- & -- \\ 
Conv1D Transpose & 64 & 8 & Adam & MSE \\ 
BatchNormalization & -- & -- & -- & -- \\ 
Conv1D Transpose & 1 & 8 & Adam & MSE 
\end{tabular}
\caption{Model parameters} \label{tab:model-param}
\end{table}

Stacking multiple linear layers introduces hierarchical abstraction, akin to performing successive linear projections in different subspaces. It also enlarges the effective receptive field, enabling the model to capture more complex patterns across the input without relying on a single large, dense linear operator, which would be less efficient and less localized. We experimented with nonlinear activation functions (e.g., ReLU), but they performed worse than linear activation. With this choice, the model reduces to a single-layer architecture with a factorized convolution kernel. Restricting the model to linear operations makes it resemble a learned projection or compression operator, simplifies the optimization landscape, and emphasizes reconstructing essential information while avoiding unnecessary flexibility and mitigating issues such as gradient vanishing.

\begin{figure}[H]
\centering
\includegraphics[width=0.65\linewidth]{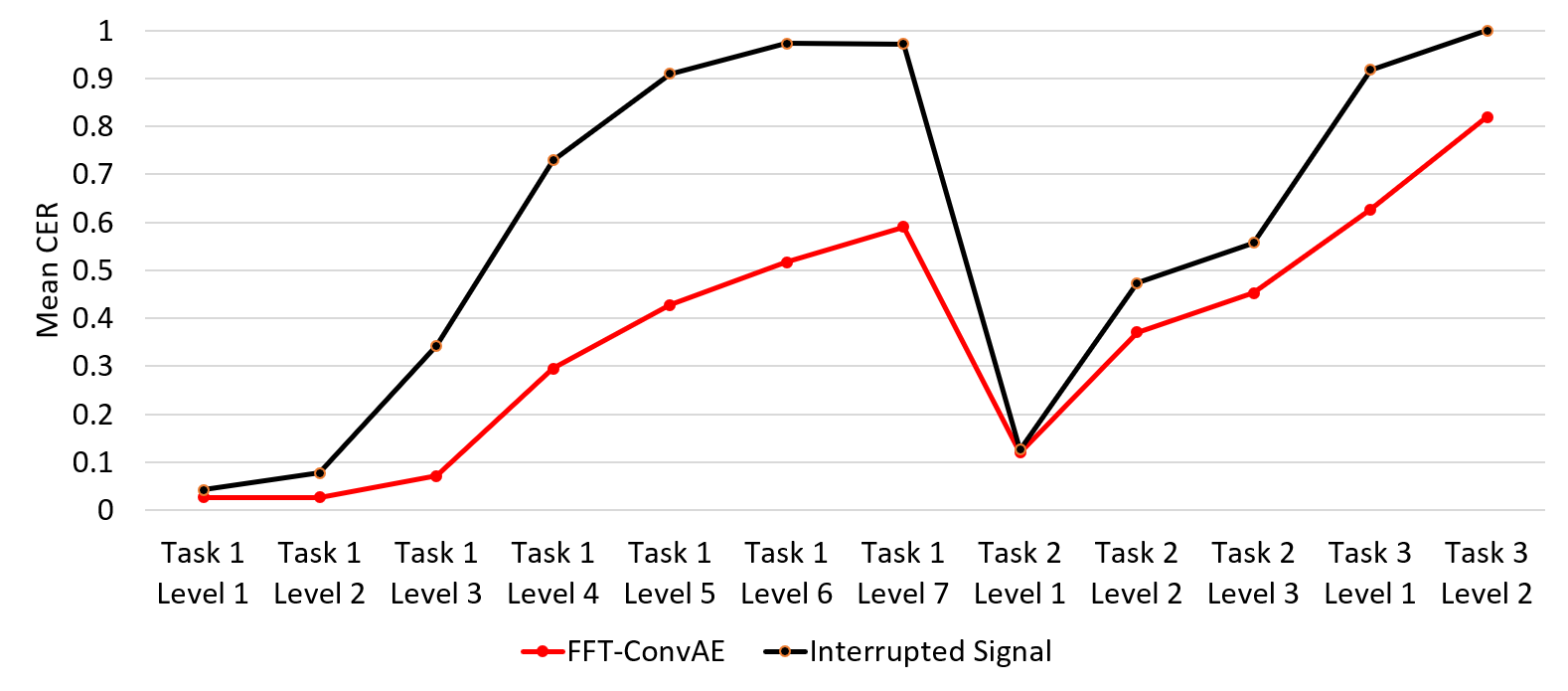}
\caption{The performance of training}
\label{fig:training-performance}
\end{figure}

\begin{figure}[H]
\centering
\includegraphics[width=.65\linewidth]{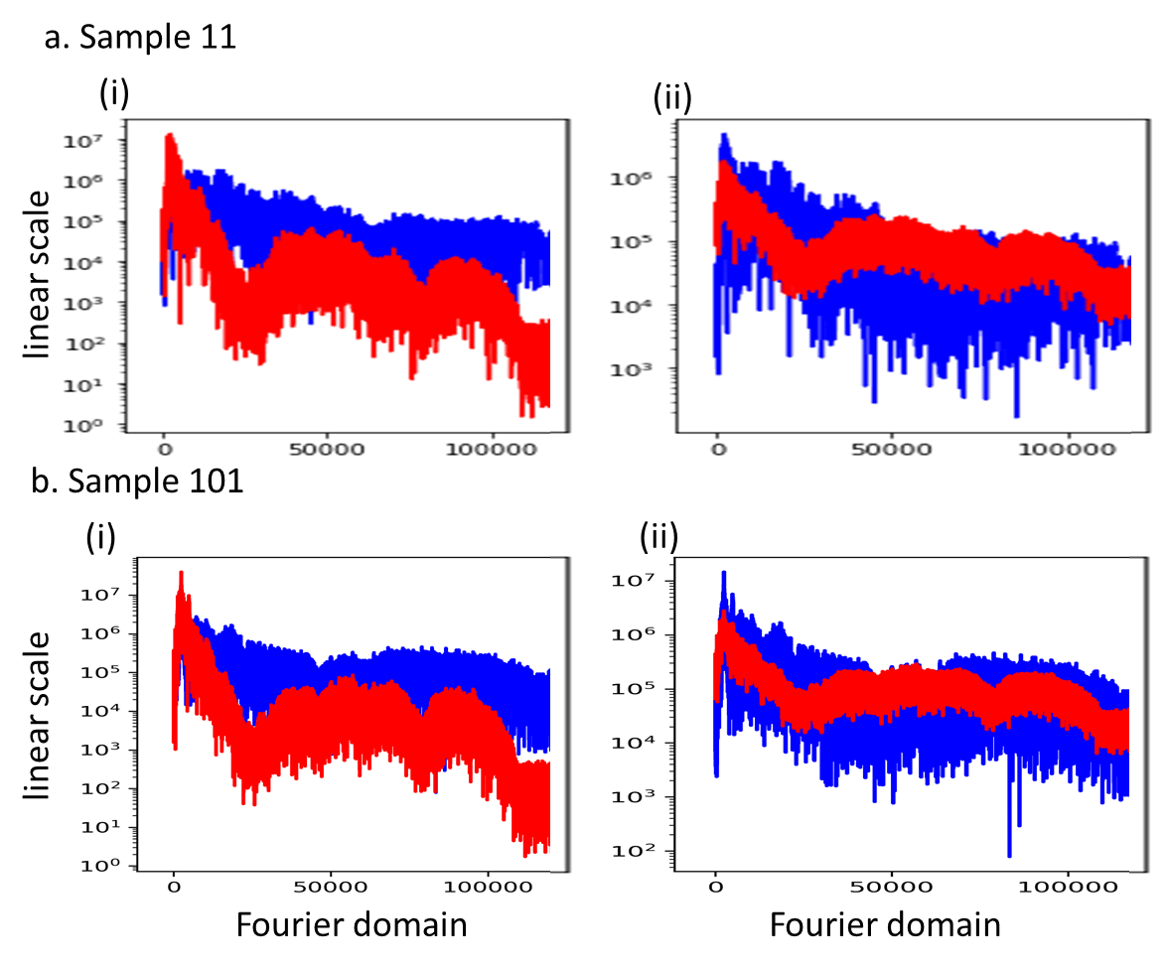}
\caption{Task~1 Level~1: Blue shows the Fourier magnitudes of the clean signal. Red indicates (i) the filtered signal and (ii) the trained signal} \label{fig:reconstruction-T1L1}
\end{figure}

\begin{figure}[H]
\centering
\includegraphics[width=.45\linewidth]{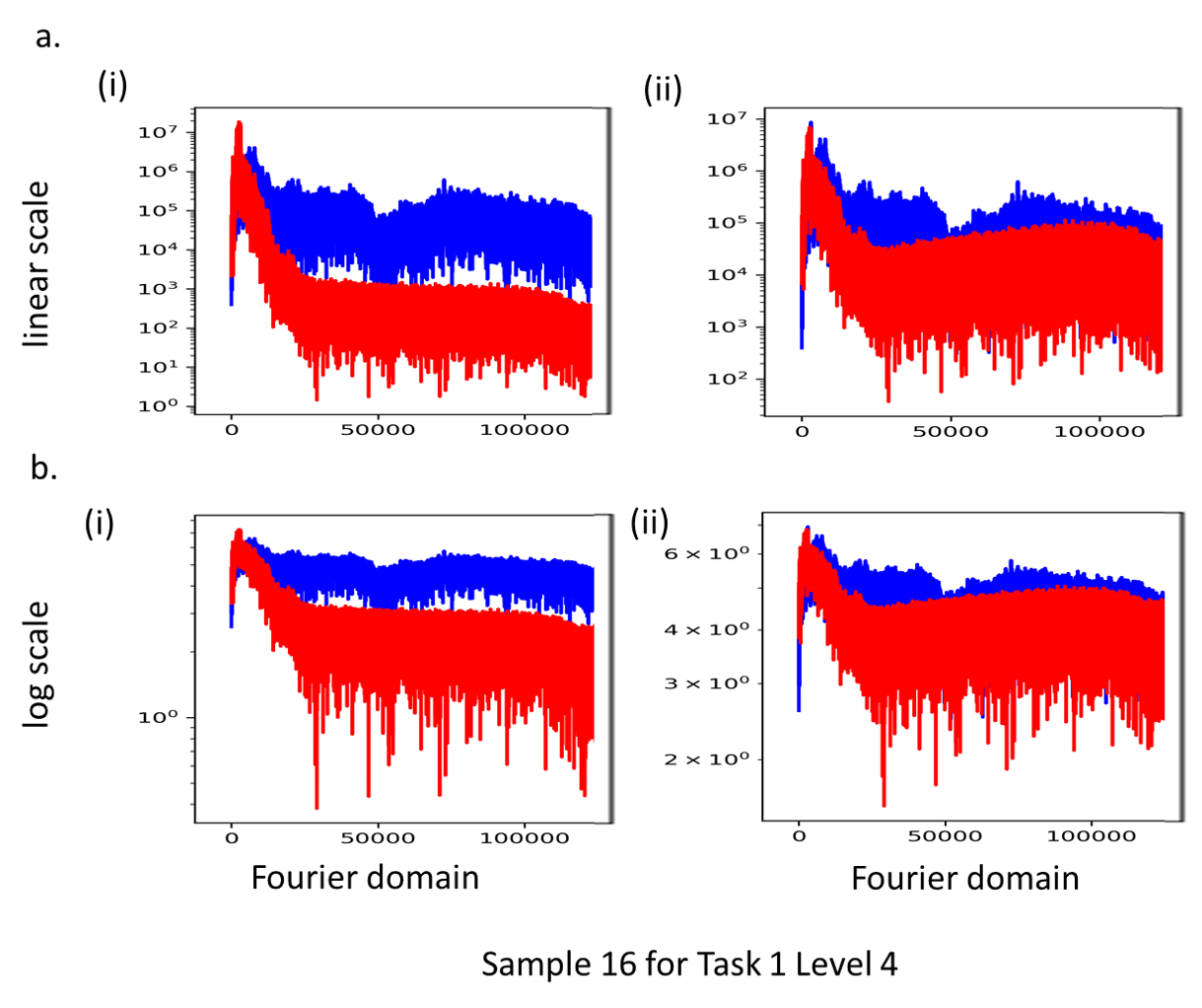}
\includegraphics[width=.45\linewidth]{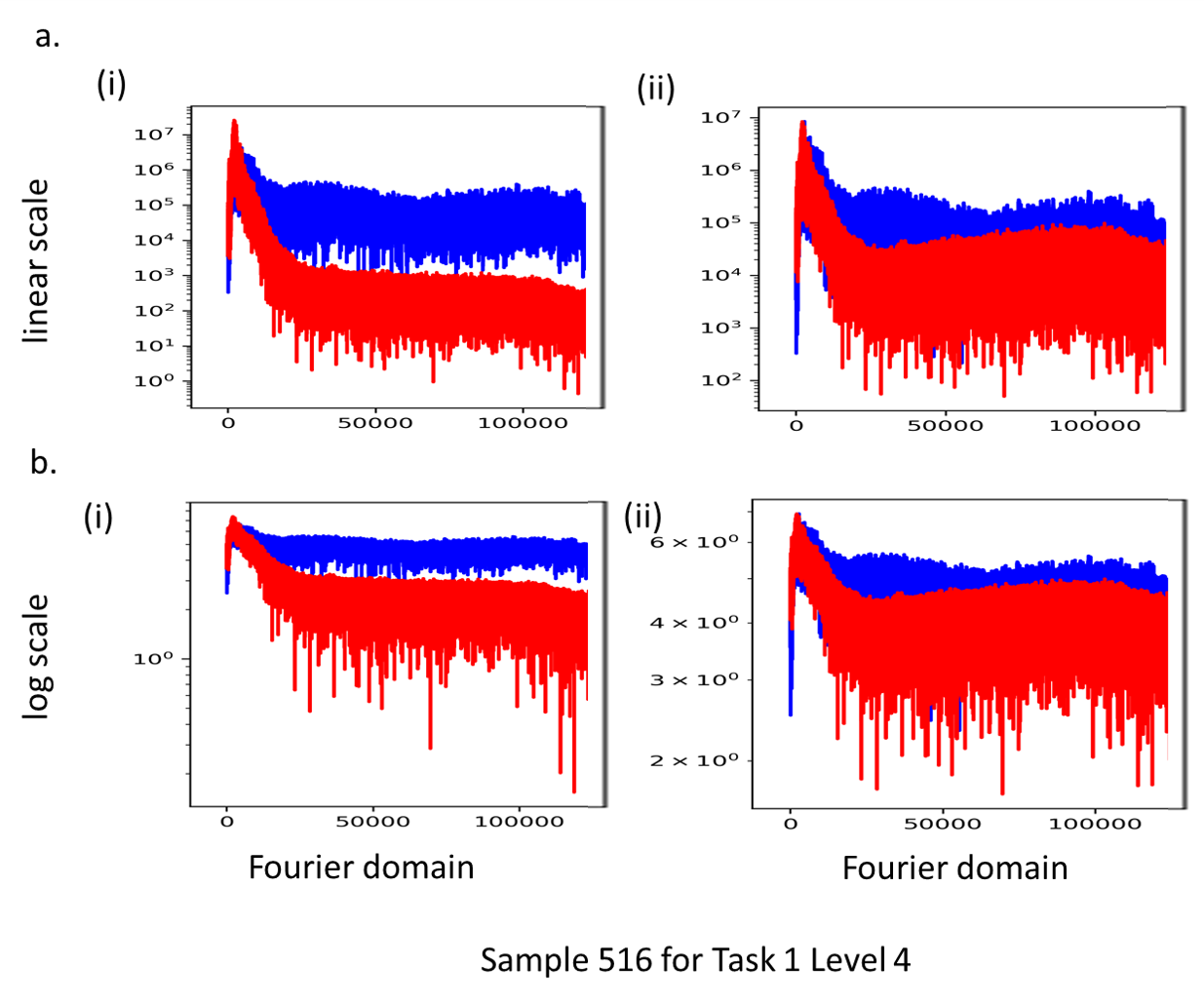}
\caption{Samples~\#16 and \#516 in Task~1 Level~4: Blue shows the Fourier magnitude of the clean signal. Red indicates (i) the Fourier magnitude of the filtered signal and (ii) the Fourier magnitude of the trained signal. }
\label{fig:T1L4-sample16-sample516}
\end{figure}

For Task~1 Levels~1--3, we take $\phi\equiv \rm Id$ (see \Cref{fig:reconstruction-T1L1} for training results in Task~1 Level~1). For all other tasks/levels, we choose $\phi(t)=\log t$. 
We employ a free, open-source Python ConvAE package from TensorFlow\footnote{\url{https://www.tensorflow.org/tutorials/generative/autoencoder}}, which is user-friendly and allows easy construction of neural networks by combining building blocks and adjusting parameters. 
Multi-scale plots are shown in \Cref{fig:T1L4-sample16-sample516}, and overall training performance, measured by CER using \texttt{evaluate.py} from the Zenodo repository\footnote{\url{https://zenodo.org/records/14007505}}, is shown in \Cref{fig:training-performance}. The $x$-axis in \Cref{fig:reconstruction-T1L1,fig:T1L4-sample16-sample516} represent the Fourier domain, i.e., the frequency ($\rm 1/s$) multiplied by a constant. For reference, the $16\,{\rm kHz}$ sample rate captures frequencies up to $8000\,{\rm Hz}$ (see \Cref{rem:bandwidth}).

We emphasize that the model itself was not trained with \texttt{evaluate.py}. As indicated in \eqref{eq:approximator}, we do not train the phase of the signal, using the dataset $\frac{ \hat{x}_{j}^{\rm deg} }{\abs{ \hat{x}_{j}^{\rm deg}} }$ and $\frac{ \hat{x}_{j}^{\rm clean} }{\abs{ \hat{x}_{j}^{\rm clean}} }$, since the model is highly sensitive to phase shifts and tends to overfit when phase training is attempted. This is further supported by the observation that the datasets $\frac{ \hat{x}_{j}^{\rm deg} }{\abs{ \hat{x}_{j}^{\rm deg}} }$ and $\frac{ \hat{x}_{j}^{\rm clean} }{\abs{ \hat{x}_{j}^{\rm clean}} }$ exhibit almost zero $R^{2}$ correlation. During testing, the dataset $\frac{ \hat{x}_{j}^{\rm clean} }{\abs{ \hat{x}_{j}^{\rm clean}} }$ is unknown, so we can only use the phase $\frac{ \hat{x}_{j}^{\rm deg} }{\abs{ \hat{x}_{j}^{\rm deg}} }$ of the degraded signal for reconstruction.

Our model is computationally lightweight, achieving real-time factors (RTF) well below 1 (see \Cref{tab:RTF}), and runs on a system with an Intel\textsuperscript{\textregistered} Core\texttrademark \, i7-10750H CPU @ $2.60\,{\rm GHz}$ (12 cores), 16GB RAM, and an 8GB NVIDIA GeForce RTX 2070 GPU. We also present the spectrograms, transcripts, and CERs of selected samples in \Cref{fig:spectrogram-T1L1,fig:spectrogram-T1L4}. \Cref{fig:spectrogram-T1L1} shows that FFT-ConvAE preserves useful signals in Sample~\#11, as the CER remains unchanged after reconstruction. Although Task~1 Level~1 is primarily a warmup level, the spectrograms of the filtered and clean signals can still differ significantly, even for high-quality audio where DeepSpeech may make errors. Spectrogram comparisons (e.g., Sample~\#11) show minimal distortion, and CER remains low, indicating that the model effectively retains both low- and high-frequency components. However, in some cases (e.g., Sample~\#101), slight over-denoising occurs, slightly increasing the CER, which highlights the model's sensitivity to fine-grained variations in clean signals. 
Nevertheless, FFT-ConvAE captures high-frequency components, enhancing overall audio quality. For Task~1 Level~4, perhaps the most relevant level, \Cref{fig:spectrogram-T1L4} demonstrates that FFT-ConvAE reduces CER in samples such as \#16 and \#516, highlighting the model's ability to effectively learn high-frequency information from the clean signal. The spectrograms show that FFT-ConvAE is able to restore missing energy in higher frequencies, illustrating the benefit of learning magnitude patterns in the Fourier domain. 

\begin{table}[H] 
\rowcolors{2}{lightgray!50}{white}
\renewcommand{\arraystretch}{1.5}  \centering
\begin{tabular}{c|c|c|c}
 & processing time (seconds) & audio length (seconds) & real-time factor\\
\hline\hline
Task 1 Level 1 & 73 & 2400 & 0.03\\
Task 1 Level 2 & 93 & 2440 & 0.04\\
Task 1 Level 3 & 63 & 2444 & 0.03\\
Task 1 Level 4 & 123 & 2444 & 0.05\\
Task 1 Level 5 & 63 & 2444 & 0.03\\
Task 1 Level 6 & 93 & 2444 & 0.04\\
Task 1 Level 7 & 73 & 2444 & 0.03\\
Task 2 Level 1 & 53 & 1292 & 0.04\\
Task 2 Level 2 & 63 & 1120 & 0.06\\
Task 2 Level 3 & 53 & 1184 & 0.04\\
Task 3 Level 1 & 53 & 1120 & 0.05\\
Task 3 Level 2 & 63 & 1120 & 0.06 
\end{tabular}
\caption{Real-time factor (RTF)} \label{tab:RTF}
\end{table}

\begin{figure}[H]
\centering
\includegraphics[width=.9\linewidth]{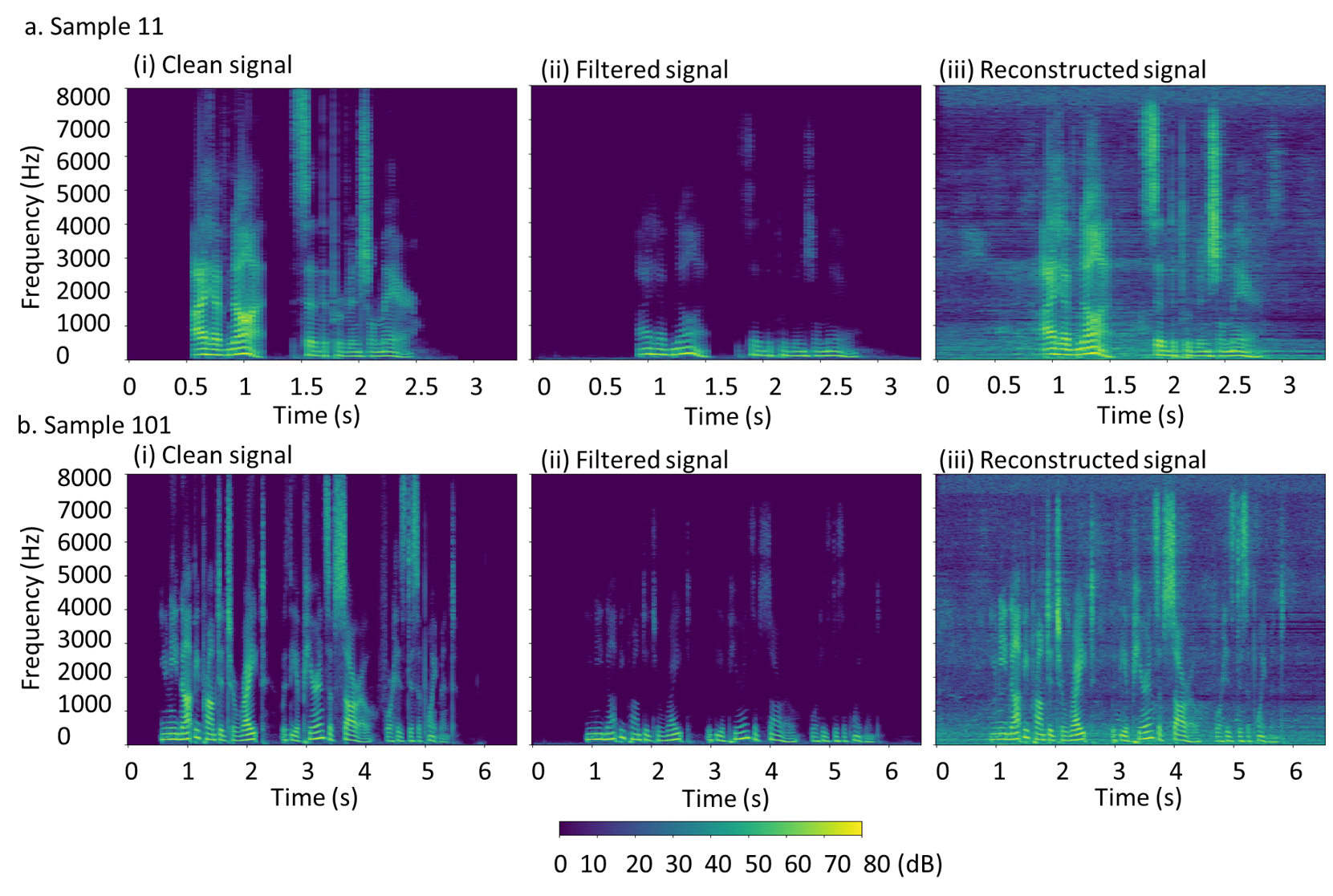}

\renewcommand{\arraystretch}{1.5} \centering
\begin{tabular}{p{2.3cm}|p{4cm} p{4cm} p{4cm}}
 & Before reconstruction & After reconstruction & True text \\
 \hline \hline
 & (${\rm CER}=0$) & (${\rm CER}=0$) & \\ 
Sample~\#11 & i have not said the provincial mayor &  i have not said the provincial mayor & I have not, said the Provincial Mayor \\ 
 \hline 
 & (${\rm CER}=0$) & (\tre{${\rm CER}=0.0694$}) & \\ 
Sample~\#101 & You need not be prompted to write with the appearance of sorrow for his disappointment. &  you need not be prompted to write \tre{that} the appearance of sorrow or his disappointment & You need not be prompted to write with the appearance of sorrow for his disappointment
\end{tabular}

\caption{Spectrogram, texts transcribed by \texttt{evaluate.py} and CER of (a) Sample \# 11 and (b) Sample \# 101 in Task~1 Level~1}
\label{fig:spectrogram-T1L1}
\end{figure}

\begin{figure}[H]
\centering
\includegraphics[width=.9\linewidth]{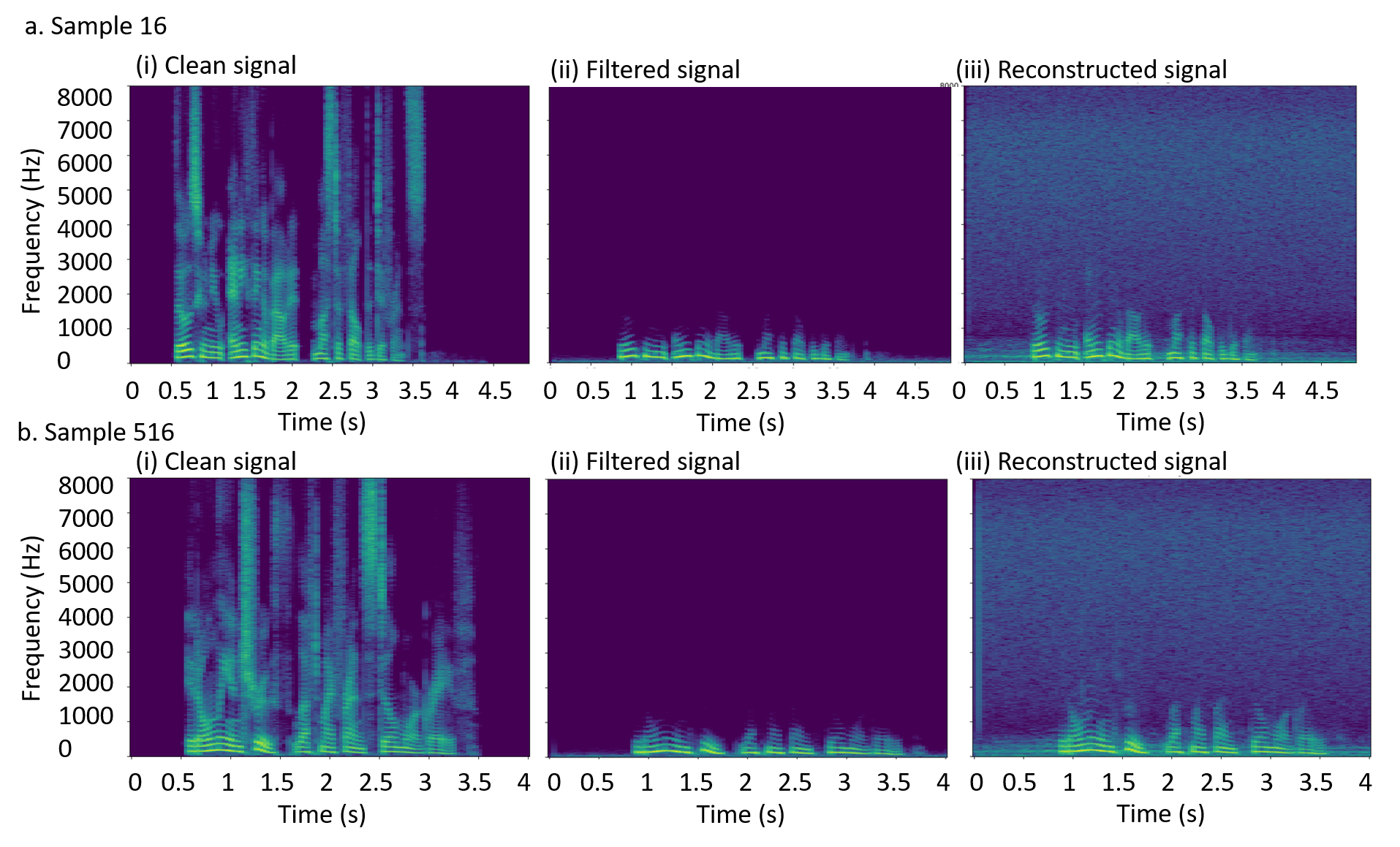}

\renewcommand{\arraystretch}{1.5} \centering
\begin{tabular}{p{2.3cm}|p{4cm} p{4cm} p{4cm}}
 & Before reconstruction & After reconstruction & True text \\
 \hline \hline 
 & (${\rm CER}=0.5$) & (${\rm CER}=0.115$) & \\
Sample~\#16 & onn about a mateself the difference &  those e ye anything about it must have felt the difference & Those who knew any thing about it, must have felt the difference \\ 
 \hline 
 & (${\rm CER}=0.436$) & (${\rm CER}=0.128$) & \\ 
Sample~\#516 &  noman fhop left my sond still more grose &  et only inpruthd lest my sriend still more grave & It only, in truth, left my friend still more grave
\end{tabular}

\caption{Spectrogram, texts transcribed by \texttt{evaluate.py} and CER of (a) Sample \# 16 and (b) Sample \# 516 in Task~1 Level~4}
\label{fig:spectrogram-T1L4}
\end{figure}

\newpage

\section{Results\label{sec:results}} 

\begin{figure}[H]
\centering
\includegraphics[width=.8\linewidth]{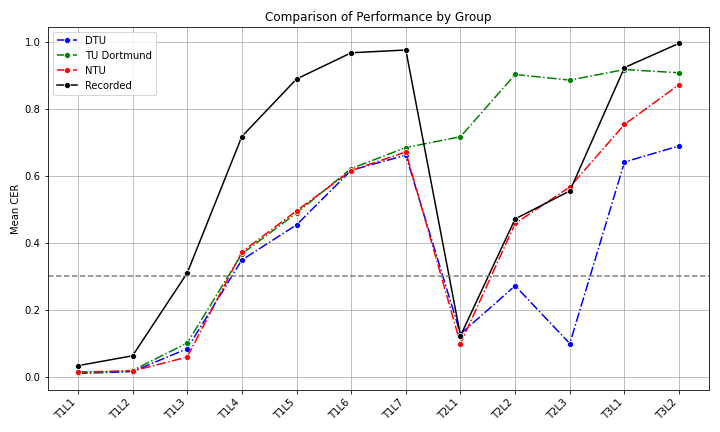}
\caption{Our group wins the second place -- labeled as \tre{NTU}} 
\label{fig:HSC2024results}
\end{figure}

The average CER, as evaluated by the organizers, is shown in \Cref{fig:HSC2024results} and is available on the official Helsinki Speech Challenge 2024 results page\footnote{\url{https://blogs.helsinki.fi/helsinki-speech-challenge/results/}}.

We next compare the performance of FFT-ConvAE during training (see \Cref{fig:training-performance}) and testing (see \Cref{fig:HSC2024results}). Performance remains consistent across all Task~1 levels in both stages. However, the mean CER increases for Tasks~2 and~3 during testing, likely due to high-frequency components in the phase of the Fourier transform, since the model directly relies on the phase of the degraded signal. 

Beyond the quantitative metrics, our observations of the reconstructed audio spectrograms reveal important qualitative trends. The choice in \eqref{eq:approximator} reflects our goal of training only stable features while keeping the model lightweight. This approach performs well for Task~1 (filtering experiments), as evidenced by comparisons with other groups' results. For T1L1--T1L3, where the audio is already of high quality, FFT-ConvAE (with $\phi \equiv \rm Id$) largely preserves the signal. At higher Task~1 levels, where the audio is more corrupted, the model (with $\phi(t) = \log t$) still captures essential high-frequency components, improving intelligibility, as reflected in reduced CERs.
For Tasks~2 and~3, the qualitative differences are more pronounced. The reconstructed audio often contains artifacts or residual reverberations, reflecting the ill-posed nature of deconvolution and the model's reliance on the phase of the degraded signal. These observations indicate that our model is not well-suited to handle these more challenging tasks. 
Overall, the qualitative findings align with the quantitative metrics, confirming that FFT-ConvAE performs robustly on less corrupted signals while remaining computationally lightweight, but struggles as task difficulty increases.

\newpage 

\section{Discussions\label{sec:discussions}} 

First, we explain why we use the Fast Fourier Transform (FFT) instead of the more commonly used Short-Time Fourier Transform (STFT) in audio applications. The STFT can be expressed as: 
\begin{equation*}
\tilde{x}_{k} = \sum_{m=-N}^{N-1}w_{m}x_{m+m_{0}}\exp\left(-2\pi\bfi\frac{mk}{n}\right) 
\end{equation*}
where $(x_{m_{0}-N},\cdots,x_{m_{0}+N-1})$ is a signal block of length $2N$ and $w_{m}$ is a window function that shapes the spectrum. 
In our dataset, each of the 12 real-world-like setups contains samples of similar lengths. 
Consequently, applying FFT with simple zero-padding is sufficient and convenient for training, while processing the full signal with FFT helps preserve its integrity.

Interestingly, the DTU team also proposed a simple method called the impulse response (IR) approach. They attempted to combine IR with Voicefixer, but the improvement was negligible. Remarkably, in Task~1, this naive method outperformed all sophisticated methods proposed by the top three winning teams, suggesting that the most effective strategy for Task~1 may indeed be a simple one.

\begin{figure}[H]
\centering
\includegraphics[width=.6\linewidth]{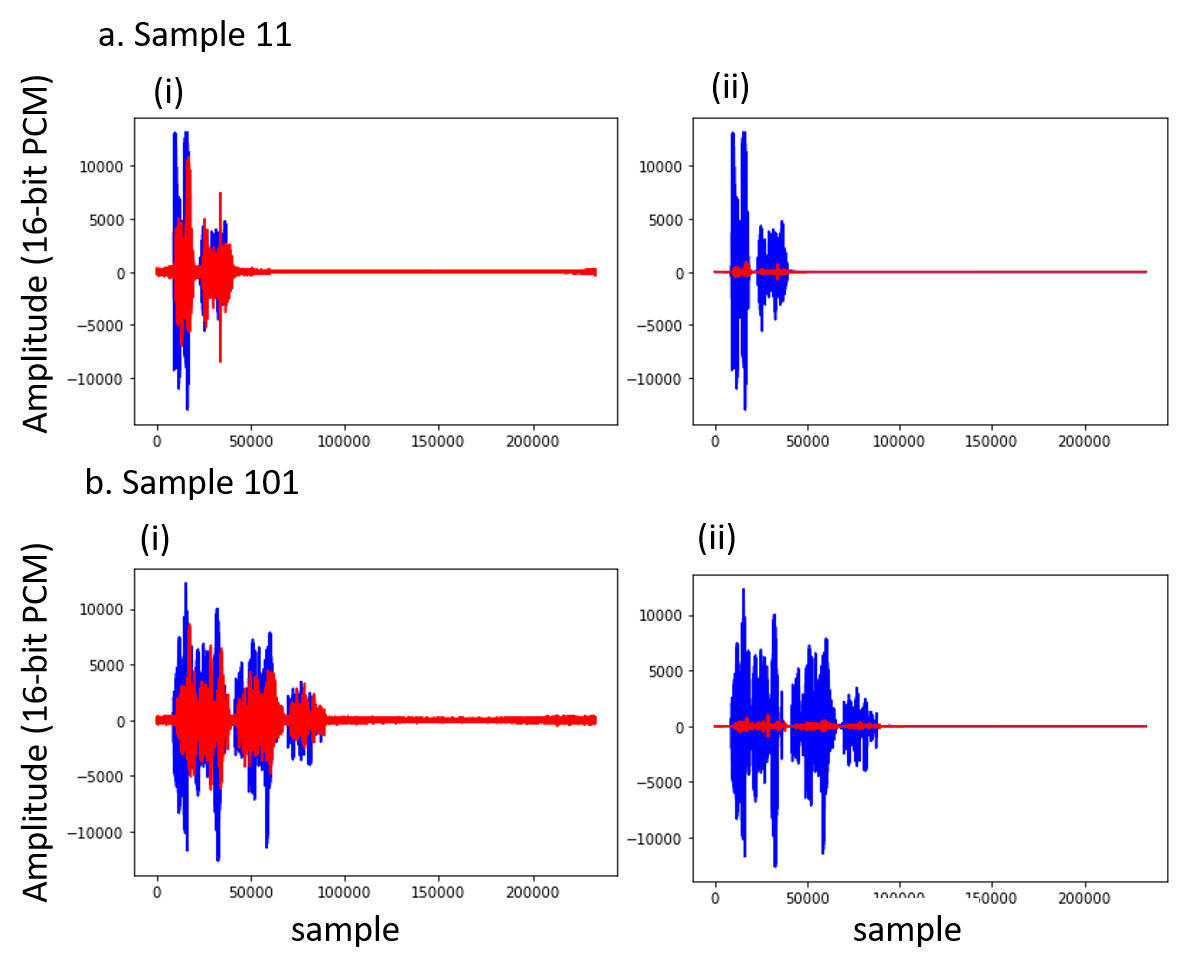}
\caption{Blue and red color represent clean and trained audio signal, respectively, by using FFT-ConvAE (left) versus pure ConvAE without using FFT (right)} \label{fig:FFT-comparison}
\end{figure}

Although the audio signal can be represented as an integer-valued vector, its highly oscillatory nature makes it difficult to handle without any suitable transform, as illustrated in \Cref{fig:FFT-comparison}. The figure shows large discrepancies between the filtered and clean signals in the original scale. After applying the Fourier transform (see \Cref{fig:FFT-comparison}(a)), these differences are noticeably reduced, and adopting a logarithmic scale on the Fourier magnitudes further minimizes the discrepancy.

\section{Conclusions\label{sec:conclusions}} 

Comparing our findings with those of the other two winning teams, it appears that the DTU team's simple impulse response (IR) method outperformed more sophisticated approaches proposed by the top three teams, suggesting that a simple strategy may be the most effective for Task~1. For Task~2, the most effective approach seems to be the DTU team's regularized impulse response (regularized IR) method, a variant of the regularization technique (see also \cite{MS12PracticalIP}). This method is neural-network-free and can be solved using interior-point or simplex methods, standard convex optimization techniques that are not typically classified as neural-network methods. Taken together, our results and those of the other teams highlight the potential of neural-network-free approaches for speech signal reconstruction, offering lightweight yet effective alternatives to the deep learning-based methods commonly used in speech enhancement tasks \cite{SpeechProcessingReview2023,FrequencyDomainMonauralSpeechEnhancement}.

\appendix

\section{Stability and Instability Mechanisms in Inverse Problems\label{sec:functional-analysis}}

This appendix is devoted to explaining certain mechanisms in inverse problems from a functional analysis perspective. In particular, we aim to clarify the notion of `features' in mathematical terms. Rather than introducing extensive functional analysis terminology, we revisit examples from \cite{KSZ2024SVDIncreasingStability} (see also \cite{KRS21InstabilityMechanism}) to illustrate the key ideas.

Given any $f\in L^{2}(\mS^{n-1})$ with $n\ge 2$, the corresponding (scaled) Herglotz wave function is formally defined by 
\begin{equation*}
A_{k}(f) := \kappa^{\frac{n-1}{2}}\left.P_{\kappa}f\right|_{B_{1}} \quad \text{with} \quad (P_{\kappa}f)(x) := \int_{\mS^{n-1}} e^{\bfi\kappa\omega\cdot x} f(\omega) \, \rmd S(\omega) \equiv (f\,\rmd S)\ehat(-\kappa x). 
\end{equation*}
By a version of Agmon-H{\"o}rmander estimate \cite[Lemma~2.3]{KSZ2024SVDIncreasingStability}, there exists a constant $C=C(n)>0$ such that for any integer $m\ge0$ one has 
\begin{equation*}
\norm{A_{\kappa}f}_{L^{2}(B_{1})} \le C(Cm\kappa)^{2m} \norm{f}_{H^{-2m}(\mS^{n-1})} \quad \text{for all $f\in L^{2}(\mS^{n-1})$,} 
\end{equation*}
where $H^{-2m}(\mS^{n-1})$ is the standard Hilbert space which can be defined in terms of the Laplace-Beltrami operator $-\Delta_{\mS^{n-1}}$ on $\mS^{n-1}$. We use Weyl asymptotics (see e.g. \cite[Theorem~8.3.1]{Tay11PDEvol1}) to simplify our quantification. The case when $m=0$ can be found in \cite[Theorem~2.1]{AgmonHormander}. This shows that 
\begin{equation}
A_{\kappa} : L^{2}(\mS^{n-1}) \rightarrow L^{2}(B_{1}) \label{eq:bounded-linear-operator-Ak}
\end{equation}
is a bounded linear operator which is compact. In addition, the analyticity of $P_{\kappa}f$ (due to Paley-Wiener-Schwartz theorem, see e.g. \cite[Theorem~10.2.1(i)]{FJ98Distribution}) implies that $f$ is uniquely determined by $A_{\kappa}f$, thus \eqref{eq:bounded-linear-operator-Ak} is injective, and it has a sequence of singular values $\sigma_{j}=\sigma_{j}(A_{\kappa})$ with $\sigma_{1}\ge\sigma_{2}\ge\cdots\rightarrow 0$, see e.g. \cite[Proposition~2.3]{KRS21InstabilityMechanism}. In order to simplified our notations, we write $A\lesssim B$ (resp. $A\gtrsim B$ or $A\simeq B$) for $A\le CB$ (resp. $A\ge C^{-1}B$ or $C^{-1}A\le B \le CA$) where $C$ is a constant independent of asymptotic parameters (here $j$ and $\kappa$). For each $\kappa\ge 1$, it was proved in \cite[Theorem~1.1]{KSZ2024SVDIncreasingStability} that the singular values $\sigma_{j}(A_{\kappa})$ of \eqref{eq:bounded-linear-operator-Ak} satisfy 
\begin{subequations} 
\begin{align}
& \sigma_{j}(A_{\kappa}) \simeq 1 && \text{for all $j\lesssim\kappa^{n-1}$,} \label{eq:stable-region1} \\ 
& \sigma_{j}(A_{\kappa}) \lesssim \exp\left(-c\kappa^{-1}j^{\frac{1}{n-1}}\right) && \text{for all $j\gtrsim\kappa^{n-1}$,} \label{eq:unstable-region1}
\end{align}
\end{subequations} 
where the constant $c>0$ and the implied constants are independent of $\kappa$ and $j$. From \eqref{eq:stable-region1}--\eqref{eq:unstable-region1}, by refining the results in \cite{KRS21InstabilityMechanism}, it was proved in \cite[Theorem~1.2]{KSZ2024SVDIncreasingStability} that a necessary condition of the existence of such a non-decreasing function $t\in\mR_{+}\mapsto\omega(t)\in\mR_{+}$ with 
\begin{equation*}
\norm{f}_{L^{2}(\mS^{n-1})} \le \omega\left(\norm{A_{\kappa}f}_{L^{2}(B_{1})}\right) \quad \text{whenever $\norm{f}_{H^{1}(\mS^{n-1})}\le 1$} 
\end{equation*}
is 
\begin{equation}
\omega(t) \gtrsim \max\left\{ t , \kappa^{-1}(1+\log(1/t))^{-1} \right\} \quad \text{for all $0<t\lesssim 1$,} \label{eq:instability-IP1}
\end{equation}
where the implied constants are independent of $\kappa$ and $t$. By inspecting the proof, one sees that the stability bound $\omega(t)\gtrsim t$ follows from \eqref{eq:stable-region1}, while the instability bound $\omega(t)\gtrsim \kappa^{-1}(1+\log(1/t))^{-1}$ follows from \eqref{eq:unstable-region1}, therefore \eqref{eq:stable-region1} and  \eqref{eq:unstable-region1} characterize the number of stable and unstable features in the inverse problem. For each fixed $\kappa>0$, from \eqref{eq:instability-IP1} we conclude that the inverse problem is ill-posed. However, can choose a large $\kappa$ to reduce the effect of the instability term $\kappa^{-1}(1+\log(1/t))^{-1}$ as well as increase the number of stable features in the sense of \eqref{eq:stable-region1}. This is called the increasing resolution phenomena. 

Similar mechanisms have been studied for the linearized inverse acoustic scattering problem \cite{KSZ2024SVDIncreasingStability}. We believe, many inverse problems, including the one in this paper, contain features that can be stably recovered, however, most features are inherently unstable to recover.

\end{sloppypar}

\bibliographystyle{custom}
\bibliography{ref}
\end{document}